# Journal of Physics Communications

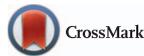

PAPER

OPEN ACCESS

# Fundamental formalism of statistical mechanics and thermodynamics of negative kinetic energy systems

Huai-Yu Wang

Department of Physics, Tsinghua University, Beijing 100084, People's Republic of China

**E-mail: wanghuaiyu@mail.tsinghua.edu.cn**

## Abstract

The solutions of a particle's Dirac equation contains a negative kinetic energy (NKE) branch. Such an energy spectrum has an upper limit but no lower limit, so that the system with this spectrum, called NKE system, is of negative temperature. Fundamental formulas of statistical mechanics and thermodynamics of NKE systems are presented. All the formulas have the same forms of those of positive kinetic energy (PKE) systems. Almost all thermodynamic quantities, except entropy and specific heat, have a contrary sign compared to those of PKE systems. Specially, pressure is negative and its microscopic mechanism is given. Entropy is always positive and Boltzmann entropy formula remains valid. The three laws of thermodynamics remain valid, as long as the thermodynamic quantities have a negative sign. Negative temperature Carnot engine can work between two negative temperatures. Since the NKE levels need not be fully filled, it is argued that the concept of Dirac's Fermion Sea can be totally abandoned.

## 1. Introduction

It is well known that, according to Dirac equation, a free particle can have two energy branches, each having corresponding eigen functions. The positive energy branch (PEB) is positive static energy (PSE) plus positive kinetic energy (PKE). Correspondingly, the negative energy branch (NEB) is necessarily negative static energy (NSE) plus negative kinetic energy (NKE). Dirac equation, as a relativistic quantum mechanics equation, is in fact applicable to all possible momentum. That is to say, it is also valid when a particle's momentum is very low. On the other hand, a particle doing low momentum motion obeys Schrödinger equation. Thus, a low momentum particle follows both Dirac equation and Schrödinger equation. This fact immediately raises a problem.

Suppose that a free particle is doing relativistic motion. It can be of either PKE or NKE. We let it, in some way, decrease its momentum gradually. When its momentum reaches a very low one, it also follows Schrödinger equation, from which there is only PKE solution, but from Dirac equation the NKE solutions should still remain. One was unable to answer whether there were the NKE solutions or not on earth. We [1] have solved this problem by proposing NKE Schrödinger equation

$$i\hbar \frac{\partial}{\partial t}\psi_{(-)} = \left(\frac{\hbar^2}{2m}\nabla^2 + V\right)\psi_{(-)} \quad (1.1)$$

which was obtained from Dirac equation by transformation $\Psi = \psi_{(-)}e^{imc^2t/\hbar}$ and under low momentum approximation. It was called NKE Schrödinger equation because there was a minus sign attached to kinetic operator. In this section, the letter $V$ denotes potential.

This equation demonstrated that a particle doing low momentum motion could be of NKE. This conclusion ought to be confirmed by experiments. Experiments were suggested to verify the existence of NKE electrons [1].

The appearance of NKE Schrödinger equation made a particle's relativistic and low momentum motions logically self consistent. The author's belief was that NKE Schrödinger equation and Schrödinger equation ought to be treated on an equal footing. The PKE solutions are just particles we have already known, while the NKE solutions are believed dark particles. Thus, personally, a NKE particle is the synonym of a dark particle. Similar to PKE particles, the properties of NKE particles should be investigated. In appendix D in the previous paper [1], we briefly outlined, by listing thirteen points, some works to be done for NKE systems. After that work, we have done some investigations: one [2] dealt with point 2, and another [3] with points 6 and 7. The present work concerns point 4.

It was pointed out [1] that for a NKE particle there could be stationary motion subject to a specific potential. This can be easily understood. We put down stationary NKE Schrödinger equation

$$\left(\frac{\hbar^2}{2m}\nabla^2 + V\right)\psi_{(-)} = E_{(-)}\psi_{(-)} \qquad (1.2)$$

and compare it to stationary Schrödinger equation

$$\left(-\frac{\hbar^2}{2m}\nabla^2 + V\right)\psi_{(+)} = E_{(+)}\psi_{(+)}. \qquad (1.3)$$

If equation (1.3) is multiplied by a minus sign, it becomes (1.2) with the potential and eigen energies taking contrary numbers, $V \to -V$ and $E_{(+)} \to -E_{(+)} = E_{(-)}$. A free particle is a simplest case of $V = 0$, and two examples were given [1]. One was an infinitely high potential barrier, and another was a Coulomb repulsive potential. Each energy spectrum had an upper limit but no lower limit. From now on, a system with this kind of energy spectrum is called NKE one. This raised a question that how the NKE particles distribute in such energy spectra. In order to answer this question, one has to resort to statistical mechanics.

This paper intends to display the statistical mechanics of NKE systems. In order for the occupation probability in any energy level not to be greater than 1, a NKE system's temperature has to be negative.

The concept of negative temperature (NT) was proposed long before [4]. There were experimental and theoretical works done for systems that could have NT. Here we merely mention some examples. Experimentally, NT was achieved in nuclear spin systems [5–11]. One common feature of these experimental systems was that their energy spectra had both upper and lower limits. Theoretically, mainly thermodynamics of possible NT systems were discussed [12–14], although statistical mechanics was also touched [12]. It was shown by an example that NT could exist naturally following Caratheodory's principle [15]. It was concluded that when a NT system was in equilibrium state, all its thermodynamic potentials, including enthalpy, free energy, Gibbs free energy and specially internal energy, reached their maxima [16, 17]. There were also arguments questioning if such a system could be in equilibrium state [18]. Numerical simulations have been done to investigate NT systems. The ultra-cold atoms in an energy band of an optical lattice could be of NT. This was because the energy band had lower and upper limits [19, 20]. Near the energy bottom (top), the atom gas was of positive (negative) temperature. The dynamics of two-dimensional NT system was simulated [21]. Recently, numerical simulation showed that a spin system could have NT [22].

The systems investigated above had two common feature. One is that their energy spectra had both upper and lower limits, and the other is that the spin systems in experiments were not isolated ones. Indeed, up to now it has been impossible to construct an isolated NKE system in labs.

In this paper, we merely discuss isolated systems and every NKE system's energy spectrum has an upper limit but no lower limit.

It has been confirmed that there is certainly dark matter in our universe. Hence, there are NKE systems in space and they are isolated. That is why it is necessary to study isolated NKE systems.

When a system's energy spectrum has a lower limit but no upper limit, this system is necessarily of positive temperature, which is the case we have already been familiar with.

When a spectrum has both upper and lower limits, this system can have both positive and negative temperature, a typical example was a spin system. Please notice that in a spin system, the distribution of spins at energy levels determines temperature. In the present work, we intend to study systems containing a great amount of molecules doing random thermal motion. For systems with both upper and lower energy limits, there are some details to be clarified, so that this kind of systems is left to discuss later.

When a system's spectrum has an upper limit but no lower limit, this system is necessarily of negative temperature, which is the case we intend to investigate in this paper.

How to determine that a system is of positive or negative temperature or both will be presented by analysis of equation (2.1) below.

After formulating the statistical mechanics, the thermodynamics of NKE systems will be presented. It will be seen that all the formulas of statistical mechanics and thermodynamics of NKE systems remain unchanged. One merely needs to replace almost all the thermodynamic quantities, except entropy and specific heat, by their negative counterparts. Nevertheless, the physical meanings of the thermodynamic quantities of PKE and NKE systems should be distinguished.

**Table 1.** Temperature ranges subordinate to definite energy spectrum ranges.

| Energy spectrum | There is no upper limit | There is an upper limit |
| --- | --- | --- |
| There is no lower limit | No such a system | $-\infty < T < 0$ |
| There is a lower limit | $0 < T < \infty$ | $-\infty < T < 0, 0 < T < \infty$ |

In section 2, the formalism of the statistical mechanics of the non-interactive dark particle systems is given. Both Bosons and Fermions are covered. An important derivation is that a NKE system produces negative pressure (NP) which was never mentioned by the above mentioned theoretical and experimental researches. In section 3, the thermodynamics of NKE systems are investigated. Section 4 is our conclusions. Dirac imagined that the NKE levels were fully occupied and they were treated as antiparticle. In appendix, it is argued that the negative energy solutions are not antiparticles.

## 2. Statistical mechanics of the non-interactive dark particle systems

Let us first consider Boltzmann distribution for distinguishable particles. The probability $\rho_n$ of a particle occupying energy level $\varepsilon_n$ is proportional to Boltzmann factor:

$$\rho_n \propto e^{-\varepsilon_n/k_B T}. \tag{2.1a}$$

The fundamental requirements are that the probability can be normalized

$$\sum_n \rho_n = 1, \tag{2.1b}$$

and at any energy level the distribution probability should be nonnegative and not greater than 1,

$$0 \leqslant \rho_n \leqslant 1. \tag{2.1c}$$

The distribution probability (2.1a) is determined by only two physical quantities: energy levels and temperature, the combinations of which that will violate (2.1b) and (2.1c) are prohibited. Let us consider three kinds of energy spectra.

The first kind of energy spectra is that each spectrum has a lower limit but no upper limit. In this case, temperature in (2.1a) has to be positive so as to guarantee equations (2.1b) and (2.1c). This is the case we have been very familiar with. PKE systems usually have this kind of spectra. The examples are a PKE free particle, a hydrogen atom and so on.

The second kind of energy spectra is that each spectrum has an upper limit but no lower limit. Since there is no lower limit, there can be negative energy levels the absolute values of which are arbitrarily large. In this case, if temperature in (2.1a) is still positive, negative and sufficiently large energy level will violate the conditions (2.1b) and (2.1c). Only negative temperature in (2.1a) can meet the conditions (2.1b) and (2.1c). This is the case we are going to study in this paper.

The third kind of energy spectra is that each has both lower and upper limits. This means that energy spectrum is in a finite range, so that energy levels are always finite. In this case the system's temperature can be either positive or negative. The examples are spin systems.

The above relationships between energy spectrum and temperature ranges are summarized in table 1.

A NKE system's energy spectrum has an upper limit but no lower limit. The simplest case is free particles solved from equation (1.2) without potential,

$$\varepsilon_{\boldsymbol{k}} = -\frac{\hbar^2}{2m}(k_x^2 + k_y^2 + k_z^2) = -\frac{\hbar^2 \boldsymbol{k}^2}{2m}. \tag{2.2}$$

Another example is that the potential in (1.2) is an infinitely high potential barrier with width $a$,

$$\varepsilon_{(-)n} = -\frac{n^2 \hbar^2 \pi^2}{2ma^2}. \tag{2.3}$$

Either energy spectrum has an upper limit but no lower limit. A system constituted by NKE particles is of negative tempurature (NT), $T < 0$.

In literature, modified equations are proposed to deal with dark matter, where a parameter called effective temperature is contained [23, 24]. We would like to have some words demonstrating that the concepts of the effective temperature in literature and of temperature in this work are totally different.

In literature, the concept of effective temperature is as follows. It appears in fundamental equation, and is believed a property of vacuum, or an intrinsic property of spacetime. The effective temperature may be positive or negative, but how to determine its range is not straightforward. It does not represent the true thermodynamic temperatures.

In this work, the concept of temperature is as follows. Temperature does not appear in the fundamental equations of quantum mechanics, and it is not a property of vacuum. It is closely related to the energy spectrum of a matter (dark matter), so that is subordinate to matter (dark matter), not to vacuum. Its range is fixed as long as the energy spectrum is known, as shown in table 1. It is a thermodynamic quantity and will appear in thermodynamic formulas in the next section. The positive temperature of a PKE system can be measured by apparatus made by PKE matter, as people have been doing. The negative temperature of a NKE system can be measured by apparatus made by NKE matter. That is the reason why we are unable to measure negative temperature directly up to now.

From a NKE spectrum and its negative temperature, we immediately see a consequence that a higher energy level has a larger occupying probability than a lower one. Especially, the highest level has the largest occupying probability, so that it is called 'negative ground state'. The levels lower than the ground state level are called 'negative excited states'. Hence, for the NKE systems, the concept of ground state, excited states and so on are just converse to those of the PKE systems.

The systems composed of identical particles obey either Bose–Einstein or Fermi–Dirac statistics.

From the statistical distribution, we are able to derive the expressions of thermodynamic quantities. We now do this for identical particle systems. In what follows we mimic the formalism for PKE systems [25].

Although the forms of the formulas in the following are the same as those of PKE system, we add, where necessary, a word 'negative' before each quantity of NKE systems to mark their differences from PKE systems.

Suppose that the states of particles are labeled by quantum number $k$. We use $\Omega$ to denote the negative thermodynamic potential of the system and it is written as $\Omega = \sum_k \Omega_k$ [25]. Then

$$\Omega_k = \mp \frac{1}{\beta} \ln(1 \pm e^{\beta(\mu-\varepsilon_k)}). \tag{2.4}$$

Hereafter, in all the formulas, the upper (lower) sign labels the Fermions (Bosons). Please keep in mind that the kinetic energy is negative, $\varepsilon_k < 0$, see (2.2), and temperature is negative $\beta = 1/k_B T$. In order to guarantee that the $\Omega_k$ is real, one has to let the chemical potential $\mu < 0$ ($\mu > 0$) for Fermion (Boson) systems. The negative chemical potential $\mu$ is a function of NT.

The mean number of particles in the state $k$ is

$$\bar{n}_k = -\frac{\partial \Omega_k}{\partial \mu} = \frac{1}{e^{\beta(\varepsilon_k-\mu)} \pm 1} = f_{\mp}(\varepsilon_k). \tag{2.5}$$

It is Fermi–Dirac (Bose–Einstein) distribution. The total particle number is calculated by $N = \sum_k \bar{n}_k$.

For a Fermion system, the zero-temperature distribution degrades to a step function,

$$f_-(\varepsilon_k) = \theta(\varepsilon_k - \varepsilon_F^0). \tag{2.6}$$

The negative chemical potential of a Fermi system at 0K is denoted by $\varepsilon_F^0$, and is termed as negative Fermi energy. At 0K, all the energies above the negative Fermi energy are occupied, which is called negative Fermi sphere, and the negative chemical potential is the lowest occupied energy. This result also confirms that for a Fermion system it must be $\mu < 0$. At finite NT, the negative chemical potential of a Fermion system is defined as the energy level at which the mean number of particles is $1/2$.

The entropy is

$$S = -\left(\frac{\partial \Omega}{\partial T}\right)_\mu = \pm k_B \sum_k \ln(1 \pm e^{\beta(\mu-\varepsilon_k)}) + \frac{1}{T} \sum_k \frac{\mu - \varepsilon_k}{e^{\beta(\varepsilon_k-\mu)} \pm 1}. \tag{2.7}$$

The above formulas are exactly the same as those of PKE systems. For a PKE system where temperature $T > 0$, its entropy is always positive. Since particles fill the energy levels from the lower ones to higher ones, $\mu > \varepsilon_k$. Therefore, both terms in equation (2.7) are positive. The same analysis applies to a NKE system where $T < 0$. Since particles fill the energy levels from the higher ones to lower ones, $\mu < \varepsilon_k$. Therefore, both terms in equation (2.7) are positive. It is easily analyzed that entropies increase with increasing NT. Entropy can also be expressed by particle number as follows.

$$S = -k_B \sum_k [n_k \ln n_k + (1 - n_k) \ln(1 - n_k)]. \tag{2.8}$$

Since entropy is always positive, we think that the Boltzmann entropy formula remains valid for NT systems: $S = k_B \ln \Theta$ where $\Theta$ is thermodynamic probability and it can be reckoned in the same way as PKE systems.

For a canonical ensemble, negative free energy is

$$F = \mp \frac{1}{\beta} \sum_k \ln(1 \pm e^{\beta(\mu - \varepsilon_k)}). \tag{2.9}$$

It is positive. The negative internal energy is

$$U = \sum_k \varepsilon_k \bar{n}_k = \sum_k \frac{\varepsilon_k}{e^{-\beta(\mu - \varepsilon_k)} \pm 1}. \tag{2.10}$$

From equations (2.7), (2.9) and (2.10) it is easily seen that

$$U = F + TS. \tag{2.11}$$

In this and next sections, the letter $V$ denotes volume. Specific heat is calculated by $C = \frac{\partial U}{\partial T}$, and the result is

$$C = \sum_k \frac{\varepsilon_k e^{-\beta(\mu - \varepsilon_k)}(\mu - \varepsilon_k)}{(e^{-\beta(\mu - \varepsilon_k)} \pm 1)^2} \frac{1}{k_B T^2} > 0. \tag{2.12}$$

The above formulas demonstrate that $\frac{N}{V} = -\frac{1}{V}\left(\frac{\partial \Omega}{\partial \mu}\right)_{T,V}$.

The $F$, $U$ and $S$ evaluated above are in a unit volume. Multiplying a volume to the negative free energy, we are able to calculate negative pressure (NP).

$$P = -\left(\frac{\partial F}{\partial V}\right)_T = \mp \frac{1}{\beta} \sum_k \ln(1 \pm e^{\beta(\mu - \varepsilon_k)}) < 0. \tag{2.13}$$

In equations (2.9)–(2.13) the formulas of a canonical ensemble are employed.

The pressure of a NKE system is always negative, i. e., dark matter generates NP. This conclusion agrees with our another work [3]. In that work, the formulas governing the motion of both PKE and NKE macroscopic bodies were derived from quantum mechanics equations. A remarkable feature was that for a NKE macroscopic body with mass $m$, the directions of its velocity $\boldsymbol{v}_{(-)}$ and momentum $\boldsymbol{p}_{(-)}$ are opposite to each other $\boldsymbol{v}_{(-)} = -\boldsymbol{p}_{(-)}/m$. This law was valid for NKE molecules although they could be regarded as microscopic particles. When molecular kinetics was applied to a NKE ideal gas, it was found that it yielded NP. Thus, the microscopic mechanism of dark matter's NP was that $\boldsymbol{v}_{(-)} = -\boldsymbol{p}_{(-)}/m$.

In the researches [4–22] mentioned in Introduction, none was able to reach the conclusion that a NT system could yield NP.

Multiplying a volume $V$ to the negative thermodynamic potential $\Omega$, we get

$$\Omega = -PV. \tag{2.14}$$

Having had the above formulas, we here explicitly write down the expressions of thermodynamic quantities of free NKE particles with the dispersion relation (2.2). The negative thermodynamic potential becomes

$$\Omega = \frac{2}{3} \frac{Vgm^{3/2}}{\sqrt{2}\pi^2 \hbar^3} |\beta|^{5/2} \int_0^\infty \frac{z^{3/2} dz}{e^{z - \beta\mu} \pm 1}. \tag{2.15}$$

The negative internal energy is

$$U = -\frac{Vgm^{3/2}}{\sqrt{2}\pi^2 \hbar^3} \int_0^\infty \frac{\varepsilon'^{3/2} d\varepsilon'}{e^{\beta(\mu + \varepsilon')} \pm 1} = -\frac{3}{2}\Omega. \tag{2.16}$$

For the dispersion relation (2.2), there stands the formula

$$PV = \frac{2}{3}U. \tag{2.17}$$

At high NT, $|\beta\mu| \ll 1$, the denominator of the integrand in equation (2.17) is expanded by Taylor series to the first order. Then negative pressure is computed to be

$$P = -\frac{gm^{3/2}}{(2\pi)^{3/2}\hbar^3 |\beta|^{5/2}} e^{\beta\mu} \sqrt{\pi}\left(1 \mp \frac{1}{2^{5/2}} e^{\beta\mu}\right). \tag{2.18}$$

The first term can be regarded as the negative pressure of distinguishable particles observing Boltzmann distribution, and the second one as the correction arising from indistinguishable effect. The correction term decreases (increases) the NP for Fermion (Boson) systems. It can be regarded as embodying the effective repulsion (attraction) effect between identical Fermions (Bosons). The less the $|T|$, the greater the repulsion (attraction) effect.

In table 2 listed are the comparisons of physical properties between particle and dark particle systems without interaction. The properties of dark systems in this table are from the results derived above. In this table, all the rows except the last four also apply to interactive systems.

**Table 2.** Comparison of physical properties between non-interactive systems of PKE particles and NKE particles.

| | Particle systems | Dark particle systems |
|---|---|---|
| Kinetic energy $K$ | $K > 0$ | $K < 0$ |
| Temperature $T$ | $T > 0$ | $T < 0$ |
| Pressure $P$ | $P > 0$ | $P < 0$ |
| Entropy $S$ | $S > 0$ | $S > 0$ |
| Specific heat $C_V$ | $C_V > 0$ | $C_V > 0$ |
| Ground state | The lowest energy level | The highest energy level |
| Excited states | In higher energy levels, the occupation probability is less. | In lower energy levels, the occupation probability is less. |
| Energy extremum principle | Energy minimum principle | Energy maximum principle |
| Grand potential $\Omega$ | $\Omega < 0$ | $\Omega > 0$ |
| Free energy $F$ | $F < 0$ | $F > 0$ |
| Internal energy $U$ | $U > 0$ | $U < 0$ |
| Chemical potential of Fermion systems $\mu$ | $\mu > 0$ | $\mu < 0$ |

Because for a NKE system, its highest energy level is the ground state energy, when a NKE system undergoes a transformation, it will always reach its highest energy as far as possible in each step of the process. Therefore the movement of the system necessarily follows energy maximum principle, which is also listed in table 2.

That the thermodynamic potentials reached their maximum in equilibrium state [16, 17, 19] was in agreement with the energy maximum principle in table 2. The thermodynamic analysis by Ramsey [12] revealed that entropy and specific heat were always positive, which was also correct.

There might be different definitions of entropy, i. e., Boltzmann entropy and Gibbs entropy, and there were arguments that which one is consistent from a mathematical and statistical point of view [26, 27]. In the present work, entropy is defined by equation (2.7), which is from statistics. There is no mathematical and statistical inconsistency. Please note that negative temperature in the present paper is for systems of NKE particles that were not explored before.

Here we investigate the statistic distributions of non-interactive NKE particles. If there are interactions between NKE particles, the energy spectrum of quasi-particles or elementary excitations should be taken into account.

## 3. Thermodynamics

Having given the statistical mechanics of NKE systems, we now turn to thermodynamics. We have known the signs of thermodynamic quantities compared to their counterparts of PKE systems, as listed in table 2. This is helpful for us to discuss thermodynamics.

First of all, the concepts of negative temperature in the sense of thermodynamic meanings should be clarified.

If the macroscopic physical quantities describing a NKE substance do not vary with time, we say that it is in equilibrium state. A system in equilibrium state has necessarily a negative temperature $T$.

The concept of NT can actually be discussed word for word as positive temperature [28], if one wants to have a deep study. Here we discuss it in view of thermal motion.

In a PKE ideal gas, its great amount of molecules does thermal motion randomly. In equilibrium, the molecule number of specific kinetic energies follows Maxwell velocity distribution, which can be derived from Boltzmann distribution. The intensity of the random thermal motion can be represented by temperature. A simple relationship is that the mean of the kinetic energy of each particle is proportional to temperature, explicitly, equals $\frac{1}{2}k_B T$ multiplied by degree of freedom.

Similarly, in a NKE ideal gas, its great amount of molecules does negative thermal motion randomly, with NKE expressed by (2.2). We say that NKE particles are doing negative thermal motion. In equilibrium, the molecule number of specific negative kinetic energies follows Maxwell velocity distribution, which can be derived from Boltzmann distribution. The intensity of the random negative thermal motion can be represented by negative temperature. A simple relationship is that the mean of the negative kinetic energy of each NKE particle is proportional to NT, explicitly, equals $\frac{1}{2}k_B T$ multiplied by degree of freedom. In our previous work [3] there was an example.

The total amount of the NKE of a NKE ideal gas is its negative internal energy. It can be calculated by statistical mechanics, such as equation (2.10). In the case of Maxwell distribution, the negative internal energy is proportional to NT. For free particles (2.2), $U = \frac{3}{2}Nk_B T$.

Suppose that there are two NKE substances A and B with NTs $T_1$ and $T_2$, respectively. If $|T_1| < |T_2|$, we say substance A is negatively colder one, or simply colder one, and B is negatively hotter one, simply hotter one. Comparatively, the former is (negatively) lower temperature, or lower NT, object and the latter is (negatively) higher temperature, or higher NT, one.

When A and B connect together, allowing negative energy exchange between them, then they will finally reach negative thermal equilibrium. That is to say, they will have the same final negative temperature $T$, $|T_1| < |T| < |T_2|$.

Apparently, in the course of approaching the equilibrium, the negative internal energy of A increases, while that of B decreases. Or the intensity of the negative thermal motion in A increases, while that in B decreases. We say that some negative heat $Q$ is transferred from B to A. Substance A absorbs negative heat and B releases negative heat. Negative heat always flows from a higher NT substance to a lower NT one.

Please note that negative heat is a reflection of NKE, while heat a reflection of PKE. In a NKE system, there is no PKE so that there is no heat. That some negative heat flows from B to A does not mean that some heat flows from A to B. The reason is that there is no positive kinetic energy, so that there is no heat in the systems.

**The zeroth law of thermodynamics** (transitive law of negative thermal equilibrium) [29].

If A and B are in negative thermal equilibrium and B and C are in negative thermal equilibrium, it is necessarily that A and C are in negative thermal equilibrium.

A NKE thermodynamic system in equilibrium is called NT thermodynamic system, and also called negative heat reservoir.

A NKE system has NP. Consequently, when its volume expands, we say that this system does negative work to outside.

In the concepts, the adjective 'negative' can be omitted, thus all the concepts are exactly the same as those of PKE system, as long as one keeps in mind that they actually means negative thermodynamic quantities.

The following statements of the three laws of thermodynamics of NT systems mimic those of PKE systems in textbook [30].

**The first law of thermodynamics**

In an arbitrary thermodynamic transformation of a NKE system, let $\triangle Q$ denote the net amount of negative heat absorbed by the system and $\triangle W$ the net amount of negative work done by the system. The first law of thermodynamics states that the quantity $\triangle U$, defined by

$$\triangle U = \triangle Q - \triangle W \tag{3.1}$$

is the same for all transformations leading from a given initial state to a given final state. It is just the change of the negative internal energy of the NKE system.

In an infinitesimal transformation, the first law reduces to the statement that the differential

$$dU = dQ - dW \tag{3.2}$$

is exact.

**The second law of thermodynamics**

*Kelvin statement*

There is no thermodynamic transformation whose sole effect is to extract a quantity of negative heat from a given negative heat reservoir and convert it entirely into negative work.

*Clausius statement*

There is no thermodynamic transformation whose sole effect is to extract a quantity of negative heat from a negatively colder heat reservoir and deliver it to a negatively hotter reservoir.

The Kelvin statement and Clausius statement are equivalent. It is easy to prove that if one statement is false, then the other is also false, and vice versa.

A NT Carnot engine is a NKE substance with volume $V$ doing the following cyclic transformation. Its initial NT is $T_1$. First, it touches a negatively heat reservoir with $T_1$ from which it absorbs negative heat $Q_1$ and its volume expands. After that it leaves the $T_1$ reservoir and does adiabatic expansion until its NT becomes $|T_2| < |T_1|$. Then it touches a negatively older reservoir with $T_2$ and does isothermal shrink releasing negatively heat $Q_2$. Finally, it leaves the reservoir and does adiabatic shrink until its NT and volume go back to $T_1$ and $V$, respectively. This cycle is called negative Carnot cycle.

The efficiency of a NT Carnot engine is

$$\eta = 1 - \frac{Q_2}{Q_1} = 1 - \frac{T_2}{T_1}. \quad (3.3)$$

It will never be greater than 1.

Clausius' theorem is still valid.

In any cyclic transformation throughout which negative temperature is defined, the following inequality holds:

$$\oint \frac{Q}{T} \leqslant 0, \quad (3.4)$$

where the integral extends over one cycle of the transformation. The equality holds if the cyclic transformation is reversible. Then, thermodynamic entropy can be defined, as Clausius did. The entropy defined in such a way never decreases.

In a reversible process, the first law can be expressed as

$$TdS = dU + PdV. \quad (3.5)$$

In this equation, temperature, internal energy and pressure are negative.

Ramsey [12] thought that according to

$$T = \frac{\partial U}{\partial S}, \quad (3.6)$$

negative temperature was arising from when entropy increased with internal energy decreased. For a NKE system, the internal energy decreasing is actually negative internal energy increasing.

The relationships between thermodynamic potentials remain unchanged, such as (2.11) and (2.14).

For negative Helmholtz free energy defined by $F = U - TS$, there is a theorem. For a mechanically isolated NKE system kept a constant NT the negative Helmholtz free energy never decreases. Its corollary is that for a mechanically isolated NKE system kept a constant NT the state of equilibrium is the state of minimum negative Helmholtz free energy. This theorem and its corollary are almost word by word copied from those for PKE system [30]. This conclusion was obtained before [16, 17, 19], and is a manifestation of energy maximum principle shown in table 2.

**The third law of thermodynamics**

The entropy of a system at absolute zero is a universal constant, which may be taken to be zero.

Here zero means temperature goes to zero from the negative side. A consequence of the third law is that a system cannot 'be negatively cooled' to zero by a finite change of the thermodynamic parameters. Here 'be negatively cooled' means that the absolute value of the system's temperature becomes smaller.

With these fundamental knowledge, one is able to investigate NKE systems' thermodynamic behaviors mimicking PKE systems.

Here we quote what Ramsey said [12]. 'It should be noted that +0 K and −0 K correspond to completely different physical states. For the former, the system is in its lowest possible energy state and for the latter it is in its highest. The system cannot become colder than +0 K since it can give up no more of its energy. It cannot become hotter than −0K because it can absorb no more energy.' The words in this paragraph were right but the last sentence was not perfect. A NKE substance cannot become negatively colder than −0 K because it can give up no more of its negative internal energy. This statement is symmetric with that of a PKE substance.

This raises a question. Is it possible for a system at −0 K to release energy since it is in its highest energy level? When one tries to answer this question, he inevitably involve the problem how a NKE system transits between its NKE energy levels. In the Universe, there is dark matter. In the author's opinion, dark matter is NKE matter. Dark particles can compose NKE systems. One NKE system can have NKE energy levels, and there can be transitions between levels under some conditions. If the transitions result in releasing energies, we may observe some, but it seems not so. This problem is left to being dealt with later, which was also mentioned in the end of appendix D of the author's previous paper [1].

# 4. Conclusions

The fundamental formalism of statistical mechanics and thermodynamics of negative kinetic energy systems are presented.

The energy spectrum of the NKE system we are discussing has an upper limit but no lower limit. From the analysis of occupying probability not greater than 1, it is determined that a NKE system must have negative temperature.

The forms of the fundamental formalism of statistical mechanics of NKE systems are exactly the same as those of PKE systems. Almost all thermodynamic quantities of NKP and PKE systems concerned in this paper have contrary signs except entropy and specific heat. Entropy and specific heat are always positive for both PKE and NKE systems.

A NKE system yields negative pressure. The microscopic mechanism is that a NKE molecule's velocity and momentum have opposite directions.

For a NKE system, the ground state, or the most stable state, is at the highest energy level. Consequently, the motion equation of a NKE systems should make the action reach maximum. Dark matter systems follow the energy maximum principle.

The four thermodynamic laws still apply to NKE, as long as an adjective 'negative' is put before the thermodynamic quantities of PKE systems where necessary. A system composed of a great amount of NKE molecules have negative thermal motion. Negative heat can flow from a negatively hotter system to a lower one. When a NKE expands, it does negative work to outside, because its pressure is negative. All thermodynamic formulas remain unchanged.

This work shows the good symmetry of PKE and NKE matters.

## Acknowledgments

This research is supported by the National Key Research and Development Program of China [Grant No. 2016YFB0700102].

## Appendix. The negative energy solutions of Dirac equation are not antiparticles

A negative kinetic energy system is of negative temperature, so that the occupying probability of NKE levels will not be greater than 1. The higher a NKE level, the higher the occupying probability, which is determined by distribution functions. Thus, the concept of Dirac's Fermion Sea can be totally abandoned.

At the very beginning when Dirac established his relativistic quantum equation for spin-1/2 particles, he was puzzled that a free electron having negative energy without the lower limit had seemed absurdity and nonphysical. He tried to give a reasonable explanation: 'We must therefore establish the theory of the positrons on a somewhat different footing. We assume that nearly all the negative energy states are occupied, with one electron in each state in accordance with the exclusion principle of Pauli. An unoccupied negative-energy state will now appear as something with a positive energy, since to make it disappear, i. e., to fill it up, we should have to add to it an electron with negative energy. We assume that these unoccupied negative energy states are the positrons. These assumptions require there to be a distribution of electrons of infinite density everywhere in the world. A perfect vacuum is a region where all the states of positive energy are unoccupied and all those of negative energy are occupied.' [31].

Dirac thought that all the NKE levels were filled and the NKE solutions represented antiparticles. In the following, we show that this idea will cause contradictions.

1. Dirac equation for a free electron is

$$i\hbar \frac{\partial}{\partial t} \Psi = (-c\boldsymbol{\alpha} \cdot i\hbar \nabla + mc^2 \beta)\Psi. \tag{A.1}$$

According to Dirac's idea, the NKE level should be fulfilled forming an electron sea. On the other hand, a free positron also obeys (A.1). If its NKE levels are fully filled, we ought to have a positron sea in vacuum, and a hole in it is an electron. This is an obvious contradiction.

2. We make transformation

$$\Psi = \psi_{(+)} e^{-imc^2 t/\hbar}. \tag{A.2}$$

Then the low momentum approximation of (A.1) is Schrödinger equation

$$i\hbar \frac{\partial}{\partial t} \psi_{(+)} = -\frac{\hbar^2}{2m} \nabla^2 \psi_{(+)}. \tag{A.3}$$

There are NKE solutions in (A.1), but not in (A.2). In fact, we have shown [1] that the transformation

$$\Psi = \psi_{(-)} e^{imc^2 t/\hbar} \tag{A.4}$$

resulted in NKE Schrödinger equation

$$i\hbar \frac{\partial}{\partial t}\psi_{(-)} = \frac{\hbar^2}{2m}\nabla^2 \psi_{(-)} \tag{A.5}$$

which is low momentum approximation of the NKE solutions of (A.1). However, no one thinks (A.5) is the equation for an antiparticle.

3. We have solved Klein's paradox [2]. The key is to piecewise solve Dirac equation in such a way that in the region where the particle's energy $E$ is greater (less) than the potential $V$, the solution of the positive (negative) energy branch is adopted. Suppose that a free electron meets a potential barrier $V$ being greater than the electron's energy. Then, after it enters the barrier, the NKE solution should be adopted. If the NKE solution represents its antiparticle, this means that as soon as an electron enters a potential barrier, it becomes a positron. That is to say, before and after the barrier, the electron has different charges. This is impossible.

4. Equation (A.1) has PKE and NKE solutions. When a particle is subject to a potential $V$, Dirac equation becomes

$$i\hbar \frac{\partial}{\partial t}\Psi = (-c\boldsymbol{\alpha} \cdot i\hbar \nabla + V + mc^2\beta)\Psi. \tag{A.6}$$

It has also PKE and NKE solutions. If the NKE levels are fully occupied forming a NKE sea, we see that each potential $V$ will cause a NKE, so that there are infinite such seas. Moreover, if a potential $V$ is generated in some lab, the corresponding NKE is also generated, but before the $V$ is generated, there was no such a sea. This ridiculous.

5. Suppose that a particle carries an electric charge $q$ and is subject to an electromagnetic potential $(\boldsymbol{A}, \phi)$. Dirac equation is

$$i\hbar \frac{\partial}{\partial t}\Psi = (c\boldsymbol{\alpha} \cdot (-i\hbar \nabla - q\boldsymbol{A}) + q\phi + mc^2\beta)\Psi. \tag{A.7}$$

If this equation's NKE solution represent an antiparticle, the charge of the antiparticle should be $-q$. However, the charge has been assigned in equation (A.7). It is impossible to solve two particles with contrary charges from one equation. In fact, the antiparticle with charge $-q$ observes Dirac equation

$$i\hbar \frac{\partial}{\partial t}\Psi = (c\boldsymbol{\alpha} \cdot (-i\hbar \nabla + q\boldsymbol{A}) - q\phi + mc^2\beta)\Psi. \tag{A.8}$$

Obviously, a particle with a charge $q$ and an its antiparticle, when subject to an electromagnetic potential $(\boldsymbol{A}, \phi)$, obey different equations.

6. By transformations (A.2) and (A.4), the low momentum approximations of (A.7) and (A.8) can be derived. The low momentum approximations of the PKE and NKE solutions of (A.7) are respectively

$$i\hbar \frac{\partial}{\partial t}\psi_{(+)} = \left(\frac{1}{2m}(-i\hbar \nabla - q\boldsymbol{A})^2 + q\phi\right)\psi_{(+)} \tag{A.9}$$

and

$$i\hbar \frac{\partial}{\partial t}\psi_{(-)} = \left(-\frac{1}{2m}(-i\hbar \nabla - q\boldsymbol{A})^2 + q\phi\right)\psi_{(-)}. \tag{A.10}$$

The low momentum approximations of the PKE and NKE solutions of (A.8) are respectively

$$i\hbar \frac{\partial}{\partial t}\psi_{(+)} = \left(\frac{1}{2m}(-i\hbar \nabla + q\boldsymbol{A})^2 - q\phi\right)\psi_{(+)} \tag{A.11}$$

and

$$i\hbar \frac{\partial}{\partial t}\psi_{(-)} = \left(-\frac{1}{2m}(-i\hbar \nabla + q\boldsymbol{A})^2 - q\phi\right)\psi_{(-)}. \tag{A.12}$$

If a NKE solution means an antiparticle, equations (A.9) and (A.10) should be regarded as the equations of a pair of electron and positron. However, no one thinks so. Instead, equations (A.9) and (A.11) are the equations of a pair of electron and positron.

7. Equation (A.1) has four solutions, denoted by $\psi_{(+)\uparrow}$, $\psi_{(+)\downarrow}$, $\psi_{(-)\uparrow}$ and $\psi_{(-)\downarrow}$, respectively. The former two belong to positive energy $E_{(+)}$ and the latter two to negative energy $E_{(-)}$. The expressions of the solutions are equations (A.16)–(A.19) in our previous paper [1]. Let us inspect, for instance, the solution $\psi_{(+)\uparrow}$. Although this is an eigen function of $E_{(+)}$, it contains two major and two secondary components, the latter belonging to $E_{(-)}$. That is to say, a PKE solution actually contains NKE components. We cannot say that this solution contains its antiparticle components. If the particle has electric charge and NKE components are explained as antiparticles, each eigen function will be the mix of components with contrary charges. Furthermore, the greater the particle's momentum, the larger the proportion of the secondary components. One cannot say that the antiparticle ingredients vary with momentum. As momentum goes to infinity, the proportion of major and secondary components will tend to 1∶1. If the solution is the mix of components with contrary charges, the total charge seems to tend to be zero.

8. If the NKE components contained in an electron's PKE solution are interpreted as its antiparticle, its PKE components are electron and NKE ones to positron. When it is interacted by a proton, the PKE components feel an attraction, while the NKE components should feel a repulsive repulsion from the proton. However, people never treat an electron in this way.

9. If the matrix element of an interaction Hamiltonian between the PKE and NKE solutions is not zero, it cannot be regarded as the transition between the states of two different charges, or charge will not conserve.

People may think that Dirac, after all, based on his concept of Fermion Sea, correctly foretold positrons and later confirmed by experiments, which indicated that the hole theory seemed right. I hereby list four famous examples that correct conclusions were drawn from wrong premises.

1). Carnot's theorem that started the engine theory had been proved on the basis of caloric theory.

The fact was that there was no caloric such kind of matter.

2). Maxwell derived equations of electromagnetic fields by Either model.

The fact was that there was no Either.

3). Lorentz transformation in special relativity had firstly been derived from imagined properties of electrons.

The fact was that for electrons there were no such properties imagined by Lorentz. Later, Einstein gave correct derivation of Lorentz transformation.

4). Dirac predicted positrons by envisaged electron sea where all the negative energy levels were filled by electrons.

The fact was that there is no need to fully fill all the negative energy levels.

At last, we mention one fact by which one may have a question. In quantum electrodynamics and quantum field theory, calculations have been done by assuming the NKE solutions of Dirac equation as antiparticles, and a great amount of calculated results have been almost perfectly in agreement with experiments. However, here we stress that the NKE should not be regarded as antiparticles. Is there any contradiction? This will be made clear later.

## ORCID iDs

Huai-Yu Wang  https://orcid.org/0000-0001-9107-6120